# Initiation of nuclear reactions under laser irradiation of Au nanoparticles in the presence of Thorium aqua-ions

A.V. Simakin and G.A. Shafeev

Wave Research Center of A.M. Prokhorov General Physics Institute of the Russian Academy of Sciences, 38, Vavilov street, 119991 Moscow Russian Federation

Abstract

Initiation of nuclear reactions in Thorium nuclei is experimentally studied under laser exposure of Au nanoparticles suspended in the aqueous solution of $Th(NO_3)_4$ ($^{232}Th$). The solutions are analyzed using either Atomic Absorption Spectrometry (AAS) or gamma-spectrometry in the range of gamma-photons energy from 0.06 to 1.5 MeV. Real-time acquisition of gamma-spectra of the probes is achieved using a portable scintillator γ-spectrometer. It is found that the reaction pathway depends in which water, either $H_2O$ or $D_2O$, the laser exposure is carried out. Laser exposure at peak intensity of $10^{13}$ W/cm$^2$ in $D_2O$ results in the decrease of probes activity of all elements of Th branching including that of $^{137}Cs$ impurity. Exposure in $H_2O$ leads to the increase of activity of elements of Th branching as well as the one of the $^{137}Cs$ impurity due to fission of Th nuclei. Saturation of the liquids ($H_2O$ or $D_2O$) with gaseous $H_2$ or $D_2$, respectively, enhances the nuclear reactions under laser exposure allowing their excitation at peak intensity as low as $10^{10}$ W/cm$^2$. Enhanced γ-activity of the probe is observed after the end of laser exposure for several hours.

Introduction

Modern lasers allow excitation of nuclear energy levels via generation of high-energy particles that appear during the interaction of laser radiation with plasma produced on a solid target. Successful excitation of nuclear levels has been reported for some isotopes of Hg and Ta under exposure of a target in vacuum to a femtosecond laser radiation. [1,2]. Emission of gamma-photons from a Ta target exposed in vacuum to peak intensity of 1018 W/cm2 in a femtosecond range of pulse duration results has been reported recently [3]. The average energy of these photons is about few MeV. Picosecond laser plasma is also a source of high-energy particles whose energy is sufficient for excitation of energy levels of nuclei in the exposed target [4].

Another possibility for laser excitation of nuclear energy levels consists in laser exposure of nanoparticles suspended in a liquid (colloidal solution). This scheme allows laser initiation of nuclear reactions, e.g., transmutation of $^{196}$Hg into $^{197}$Au [5, 6] via laser exposure of Hg nano-drops in heavy water $D_2O$. It is believed that thermal neutrons needed for this transmutation are released from Deuterium though the mechanism of this release remained unknown. The possibility to induce nuclear reactions at relatively low peak intensity of laser radiation was attributed to the local field enhancement in the vicinity of metallic nanoparticles by a factor of $10^4$-$10^5$. This may provide effective peak intensity in the liquid of about $10^{17}$ W/cm$^2$, which is already comparable with those used for exposure of solid targets in vacuum.

It is of interest to use the same approach for initiation of nuclear reactions in nanoparticles of unstable elements, such as $^{238}$U or $^{232}$Th. However, these elements are chemically reactive and would react with aqueous environment during the laser synthesis. The solution of this problem consists in using NPs of noble metals, e.g., Au, to provide the constant level of absorption in the liquid, while unstable elements can be presented in the solution as aqua-ions [7].

The aim of this work is the experimental study of possibility of laser initiation of nuclear reaction in aqueous solutions of a Thorium salt under absorption of laser radiation by Au nanoparticles. Thorium nuclei decay via the sequence of a- and b-decays as follows: $^{232}$Th → $^{228}$Ra→ $^{228}$Ac→ $^{228}$Th→ $^{224}$Ra→ etc. One may expect that thermal neutrons generated through laser exposure of Au NPs in $D_2O$ should alter the equilibrium concentration of all elements that belong to Th branching.

Experimental

Au nanoparticles (NPs) were synthesized by ablation of a bulk gold target either in $H_2O$ or $D_2O$ with the help of a Nd:YAG laser with pulse duration of 70 ns at wavelength of 1.06 μm. The details of the synthesis can be found elsewhere [8]. The resulting average size of Au NPs as determined by Transmission Electron Microscopy lies between 10 and 20 nm. The Thorium salt $Th(NO_3)_4$ was then dissolved in the colloidal solution, and the solution was divided into two parts, one of them considered as the initial solution. The second part of the solution was exposed to laser radiation. The exposure was carried out either of the as-obtained solution or under continuous purge of $H_2$ or $D_2$ for $H_2O$ and $D_2O$, respectively. The gases were obtained by electrolysis of corresponding liquids, either $H_2O$ or $D_2O$ and were supplied to the solution at atmospheric pressure.

Three laser sources were used for exposure on Au NPs in the aqueous solutions of the Th salt. These were a Nd:YAG laser, pulse duration of 30 ps, wavelength of 1.06 μm, energy/pulse of 4 mJ, reprate of 10 Hz, peak power of $10^{13}$ W/cm$^2$, a Nd:YAG laser, pulse duration of 350 ps, wavelength of 1.06 μm, energy/pulse of 350 μJ, reprate of 300 Hz, peak power of $10^{11}$ W/cm$^2$, and a Cu vapor laser, pulse duration of 10 ns, wavelength of 510/578 nm, energy/pulse of 100 μJ, reprate of 15 kHz, peak power of $10^{10}$ W/cm$^2$.

Gamma-emission from samples before and after laser exposure was characterized using a semiconductor γ-spectrometer Ortec-65195-P. This provided the analysis of sample specific activity in γ-photons from 0.06 до 1.5 MeV in Becquerel per ml. Real-time acquisition of γ-spectra of the solutions during laser exposure was achieved with the help of a portable scintillator γ-spectrometer. In the latter case the cell with the solution was fixed just on the spectrometer itself, which guaranteed the constant geometry of measurements under natural background of γ-radiation. The acquisition time was sufficiently long to provide the accuracy of measurements better than 3% in the channel with maximal number of counts indicated by the spectrometer.

Atomic Absorption Spectrometry (AAS) with inductively coupled plasma was used for measurements of concentration of all elements independently on their isotopic composition.

Results and discussions

Typical gamma-spectrum of the solution taken with a semiconductor spectrometer is shown in Fig. 1. It contains peaks of elements of Th branching that are active in emission of gamma-photons. Also it reveals the presence of $^{137}Cs$ impurity. The same spectrum taken with a portable spectrometer is shown in the inset. It has lower energy resolution, and the high resolution data from the semiconductor spectrometer were used for identification of the real-time spectra. Typical activity of samples in gamma-photons was 20-25 μR/hour at natural background of 15 μR/hour.

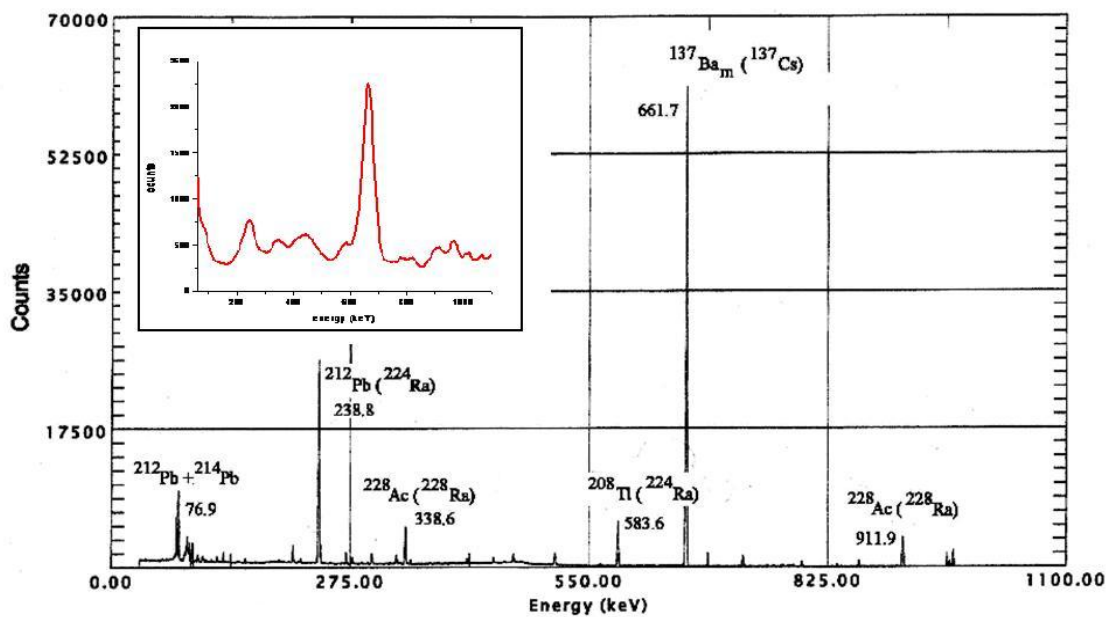

Fig. 1. Gamma-spectrum of the sample of Au NPs in the solution of $Th(NO_3)_4$ in $D_2O$. The inset in the upper left corner shows the same spectrum taken with the help of a portable scintillator spectrometer used for real-time acquisitions.

The comparison of samples subjected to picosecond laser exposure in $D_2O$ is presented in Tables 1 and 2. One can see that the activity of all elements in Th branching decreases after laser exposure, as well as that of the $^{137}Cs$ impurity. The concentration of $^{232}Th$ was calculated according to the concentration of $^{228}Ra$ using the equilibrium coefficients. This is corroborated with AAS data presented in the same Tables. AAS shows the increase of Ba concentration by a factor of 4 in case of 30 ps laser exposure. Note that exposure at $10^{13}$ W/cm$^2$ is more efficient than that at $10^{11}$ W/cm$^2$ in the sense of measurable effect. The total exposure time with 350 ps laser source was 22 hours. Therefore, laser exposure of Au NPs in $D_2O$ leads to accelerated decay of all elements of Th branching.

Table 1. Exposure in $D_2O$, Nd:YAG 1.06 μm, 30 ps, $2.88 \times 10^5$ pulses ($10^{13}$ W/cm$^2$)

| Sample | $^{228}$Ra | $^{224}$Ra | $^{232}$Th | $^{137}$Cs | $^{232}$Th, g/l | Th, g/l (AAS) | Ba, mg/l (AAS) |
|---|---|---|---|---|---|---|---|
| Initial | 39±4 | 41±5 | 39±4 | 136±9 | 9.51 | 12.7 | 0.14 |
| Exposed | 29±3 | 30±3 | 29±3 | 96±5 | 7.07 | 10.2 | 0.63 |

Table 2. Exposure in $D_2O$, Nd:YAG, 1.06 μm, 350 ps, $1.84 \times 10^7$ pulses ($10^{11}$ W/cm$^2$)

| Sample | $^{228}$Ra | $^{224}$Ra | $^{232}$Th | $^{137}$Cs | $^{232}$Th, g/l | Th, g/l (AAS) | Ba, mg/l (AAS) |
|---|---|---|---|---|---|---|---|
| Initial | 86±7 | 89±8 | 86±7 | 337±18 | 20.97 | 22.6 | 0.63 |
| Exposed | 78±7 | 78±6 | 78±7 | 299±13 | 19.02 | 23.3 | 0.83 |

The qualitatively different result is observed after laser exposure of the solution in $H_2O$ (see Table 3). Laser exposure of the solution at $10^{13}$ W/cm$^2$ results in the increase of the activity of all γ-active species. The effect is even more pronounced in $^{137}$Cs impurity. Calculations of $^{232}$Th concentration through the concentration of $^{224}$Ra give the increase of its content. However, the correct value of Th concentration is provided by AAS data.

Table 3. Exposure in $H_2O$, Nd:YAG 1.06 μm, 30 ps, $1.8 \times 10^5$ pulses ($10^{13}$ W/cm$^2$)

| Sample | $^{228}$Ra | $^{224}$Ra | $^{232}$Th | $^{137}$Cs | $^{232}$Th, g/l | Th, g/l (AAS) | Ba, mg/l (AAS) |
|---|---|---|---|---|---|---|---|
| Initial | 43±7 | 43±8 | 43±7 | 168±25 | 10.5 | 20.4 | 0.30 |
| Exposed | 61±9 | 63±10 | 61±9 | 251±38 | 14.9 | 18.6 | 0.41 |

According to them, the Th content decreases after laser exposure. Laser exposure shifts the equilibrium in Th branching, so the equilibrium ratios of concentrations in this branching are no more valid. $^{137}$Cs does not belong to Th branching, it is a fission fragment of $^{232}$Th with Y and Sr being other fragments. The increase of $^{137}$Cs concentration indicates to the fission of $^{232}$Th nuclei upon laser exposure in $H_2O$. No measurable effect of laser irradiation of Au NPs in $H_2O$ was observed at $10^{11}$ W/cm$^2$ (350 ps) even at elevated exposure time.

The experiments described above were carried out using the as-prepared colloidal solution. Saturation of the liquid with $H_2/D_2$ enhances the rate of nuclear transformations and allows its observation under much shorter exposures.

Fig. 2, a shows the difference γ-spectrum of the sample of Au NPs in $D_2O$ with Th salt under laser exposure and that of the initial sample. The peak indicated by arrow corresponds to the enhanced γ-activity of $^{232}$Ac, which is the nearest element to $^{232}$Th in its α-decay. This is coherent with previous observations of enhanced decay of $^{232}$Th under exposure in $D_2O$ (see Table 2).

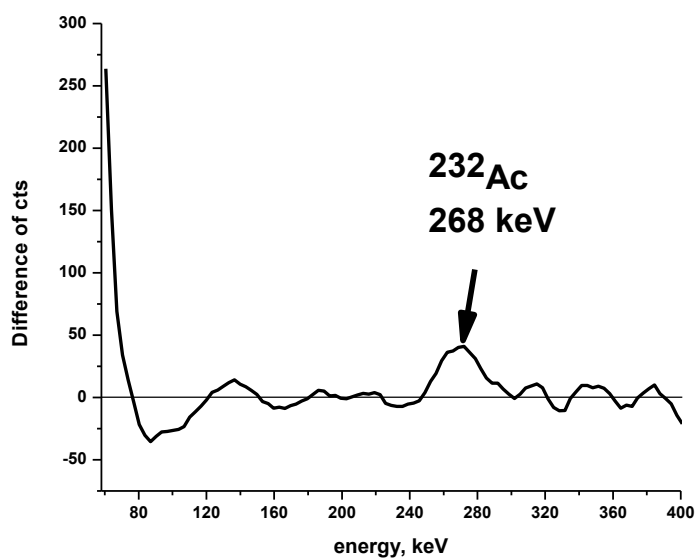

Fig. 2, a. Difference spectrum of the sample of Au NPs exposed in $D_2O$ with $Th(NO_3)_4$ to a 350 ps laser radiation and that of the sample before laser irradiation. Acquisition time was of 6 hours for both samples.

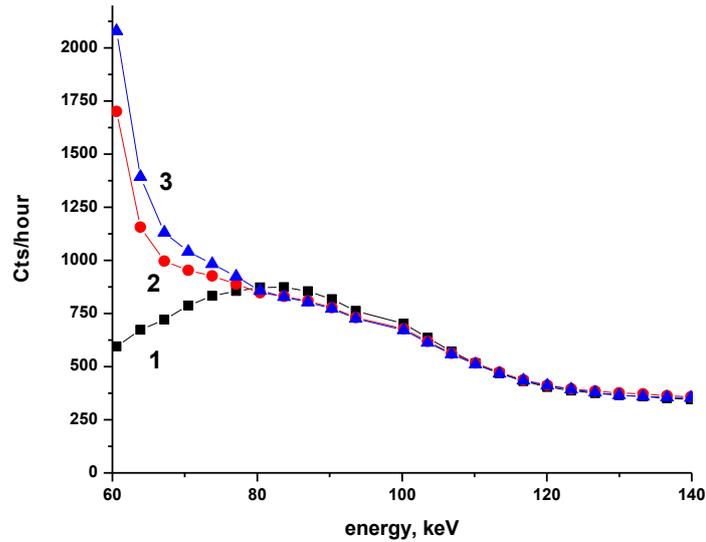

Fig. 2, b. Low-energy region of γ-spectrum of the colloidal solution of Au NPs with Thorium salt in $D_2O$ purged with gaseous $D_2$ taken before laser exposure (1), in real time (2), and after laser exposure (3). Nd:YAG laser, pulse duration of 350 ps, peak power of $10^{11}$ W/cm$^2$. Acquisition time for each of three spectra was 6 hours.

Also, large increase of gamma-activity is observed in the low-energy part of spectrum. The details of this part of spectrum are presented in Fig. 2, b. The peaks are not resolved but this energy corresponds to emission by $^{212}$Pb and $^{214}$Pb visible in Fig. 1. The striking feature is the increased γ-activity of the sample compared to the initial sample after the end of laser exposure (curve 3). The activity comes back only several hours later. So, purging of gaseous $D_2$ through the colloidal solution of Au NPs allows observations of significant difference in γ-activity of the samples under much shorter exposure times.

Purging of gas through the liquid with suspended Au NPs allows observations of nuclear reactions at even lower peak intensity of laser radiation. Fig. 3 shows the low-energy part of the g-spectrum of solution exposed to radiation of a Cu-vapor laser under purge of $H_2$ through $H_2O$. Again, the enhanced γ-activity of the sample is observed after the end of laser exposure (the laser was still on but the beam was shut down). The activity of the sample goes down only the next day several hours later.

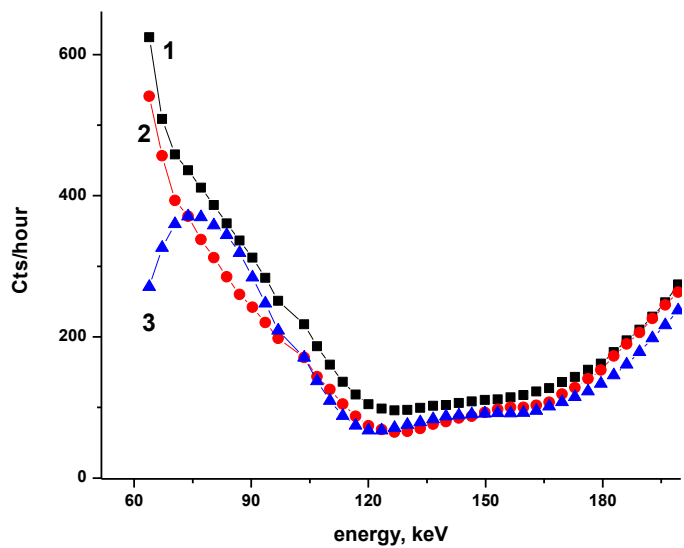

Fig.3. Low-energy region of γ-spectrum of the colloidal solution of Au NPs with Thorium salt in $H_2O$ purged with gaseous $H_2$ taken in real time (1), after the laser exposure (2), and next day (3). A Cu-vapor laser, peak power of $10^{10}$ W/cm$^2$. Acquisition time for each of three spectra was 4 hours.

Different reaction pathways observed under exposure in $H_2O$ and $D_2O$ imply different interaction of these compounds with Au NPs. This interaction is not related to chemical one since chemical properties of these two waters are the same. Indeed, NPs are molten during their synthesis by laser ablation. Accordingly, the water vapor around the NPs can be partially dissociated. Previous observations show that the emission of the plasma produced by laser beam at intensity of $10^{11}$ W/cm$^2$ in a colloidal solution of Au NPs contains the emission line of atomic Au. Its excitation requires at least 5 eV, so about 1/3 of $H_2O/D_2O$ molecules around the nanoparticle are dissociated into $H_2/D_2$ and $O_2$ (corresponding dissociation energy is of 12.6 eV). $H_2/D_2$ dissolve in the metal, while the solubility of O is much lower than that of H/D due to larger size. This process is very efficient in view of high specific surface of Au NPs used in this work since their surface is as high as 10 m$^2$ per 1 ml of colloidal solution. Saturation of the liquid with $H_2/D_2$ increases the quantity of these gases in Au NPs. If the solidification rate of NPs is sufficiently high, then the dissolved gases remain inside the NPs.

Further interpretation of the present results requires additional hypothesis and cannot be explained on the basis of known facts. One might suggest the following. NPs oversaturated with $H_2/D_2$ are heated by subsequent laser pulses. As the result, they are compressed by expanding vapors of the liquid around them. Note that NPs are optically thin, so the compression is symmetric. This automatic confinement of the pressure around gas-filled NPs provides some nuclear particles that are capable of inducing nuclear reaction in the Th nuclei presented in the solution. Most probably, these particles are neutrons, and the energy distribution of these neutrons depends on the nature of the gas that is dissolved in the NPs. According to observed nuclear transformations, exposure in $D_2O$ results in generation of thermal neutrons, while laser exposure in $H_2O$ provides more energetic neutrons capable of fission of Th nuclei.

Thus, laser exposure of Au NPs in the range of peak intensities between $10^{10}$ and $10^{13}$ W/cm$^2$ in presence of Th$^{4+}$ aqua-ions results in the initiation of nuclear transformations in $^{232}$Th nuclei. Exposure in $D_2O$ results in the decrease of activity of all elements of Th branching, as well as that of $^{137}$Cs impurity. Exposure in $H_2O$ leads to the increase of activity of the elements of Th branching along with that of $^{137}$Cs impurity due to fission of Th nuclei. The efficiency of laser exposure is enhanced if the aqueous medium, either $H_2O$ or $D_2O$ is saturated with gaseous $H_2$ or $D_2$, respectively. The peak power of laser radiation is the important parameter for initiation of nuclear transformations in the present experimental conditions. However, it can be compensated with increased number of laser pulses absorbed in the solution. The mechanism of initiation of nuclear reactions requires further studies.

Acknowledgements
The work was partially supported by Russian Foundation for Basic Research, grants ## 07-02-00757, 08-07-91950, and by Scientific School 8108.2006.2. Dr. A.V. Goulynin is thanked for gamma-measurements and helpful discussions.